# New advances in nebular photoionisation modelling

Barbara Ercolano[1]

[1]Dept. of Physics and Astronomy, University College London,
Gower St, London WC1E 6BT, UK
email: be@star.ucl.ac.uk

May 14, 2006

**Abstract**

The study of photoionised gas in planetary nebulae (PNe) has played a major role achieving, over the years, a better understanding of a number of physical processes pertinent to a broader range of fields than just PNe studies, ranging from atomic physics to stellar evolution. Whilst empirical techniques are routinely employed for the analysis of the emission line spectra of such objects, the accurate interpretation of the observational data often requires the solution of the radiative transfer (RT) problem in the nebula, via the application of a photoionisation code. A number of large-scale codes have been developed since the late sixties, using various analytical or statistical techniques mainly under the assumption of spherical symmetry and a few in 3D. These codes have been proved to be powerful and in many cases essential tools, but a clear idea of the underlying physical processes and assumptions is necessary in order to avoid reaching misleading conclusions.

The development of the codes has been driven by the observational constraints available, but also compromised by the available computer power. Modern codes are faster and more flexible, with the ultimate goal being the achievement of a description of the observations relying on the smallest number of parameters possible. In this light, recent developments have been focused on the inclusion of new atomic data, the inclusion of a realistic treatment for dust grains mixed in the ionised and photon dominated regions (PDRs) and the expansion of some codes to PDRs with the inclusion of chemical reaction networks. Furthermore the last few years have seen the development of fully 3D photoionisation codes based on the Monte Carlo method.

A brief review of the photoionisation codes currently in use is given here, with emphasis on recent developments, including the expansion of the models to the 3D domain, the identification of new observational constraints and how these can be used to extract useful information from realistic models.

# 1 Introduction

Photoionised plasma is present in many astrophysical environments, from H II regions and Planetary Nebulae (PNe), that mark the beginning and the end stages of stellar evolution, to ionised interstellar and intergalactic media, to the gas photoionised by high energy sources in AGNs and seen in distant quasars. The interpretation of their complex emission line spectra relies on our ability to disentangle the underlying physics, which involves a number of microscopic atomic processes, highly sensitive to the physical properties of the emitting gas, and to the radiation field of the ionising source(s). Whilst empirical studies of the observations are routinely carried out to extract some basic information from the spectra, the application of large-scale numerical codes is often essential to understanding these sources.

PNe, as with many ionised gas clouds, are generally transparent to the emission line radiation that is produced in their interiors. Therefore, a prediction of their spectra only requires a knowledge of the electron temperature and ionisation structure at each position in the ionised volume. This can be achieved by numerically solving the coupled ionisation balance equations and the thermal equilibrium, once the local radiation field, including the stellar and diffuse components, has been estimated by analytical or, more recently, statistical means. All major heating and cooling channels must be accounted for, as well as all ionisation and recombination processes. A comprehensive reference is the textbook by Osterbrock & Ferland (2006), additionally, a review on quantitative spectroscopy of photoionised clouds was given by Ferland (2003), whilst a summary of emission line analysis techniques can be found in a number of papers presented at a conference to honour Silvia Torres-Peimbert and Manuel Peimbert on their sixtieth birthdays (Henney et al. 2002).

The first generation of codes that were able to treat the temperature stratification in H II regions appeared in the 1960s (e.g. Flower 1968, Rubin 1968, Harrington 1968), these codes were severely hindered by the limited computing power available at the time and the lack of a comprehensive atomic data set. The fast improvements in hardware and the large volume of atomic data that has become available in the last forty years have allowed photoionisation calculations to become more and more sophisticated, boosting the predictive power of the codes and making them an essential tool for modern quantitative spectroscopic studies. At present more than eight large-scale photoionisation codes are in use. Their performance has been tested through a number of workshops, resulting in the definition of a set of rigorous benchmarks (Péquignot et al. 2001), generally referred to as the Meudon/Lexington benchmarks. The state of the art of the field at that time was described in the papers presented at the moste recent of these meeting, in *Spectroscopic Challenges of Photoionised Plasmas* (Ferland & Savin 2001). In the present paper we summarise the latest advances in nebular photoionisation modelling, highlighting the major developments that have occurred in the past five years, with particular emphasis on the applications of modern plasma codes to the modelling of PNe.

# 2 New advances

A photoionisation model aims to match all available constraints for a given target, which may include spatially resolved or unresolved optical, infrared (IR) and/or ultraviolet (UV) spectra. Images in one or more narrow-band filter are also available for many objects. A well-constrained tailored model can unveil a large volume of information regarding the nebular properties, including the gas density distribution and its chemical composition, and the properties of the ionising star(s), such as effective temperature and surface gravity. Additionally, models of PNe have been used in the past to constrain the magnitude of some then unknown atomic process rates, as in the case of the hydrogen-oxygen charge exchange process (Pequignot, Stasinska & Aldrovandi, 1978).

The desire to obtain *realistic* models for a variety of emission line sources is the main drive behind the continuous developments of large photoionisation codes in the last forty years, including the more recent advances that will be discussed here. Progress has been made on many aspects; in this paper the following five main streams will be considered: atomic data updates, the development of 3-D codes, the inclusion of dust grains, the expansion of the models to the Photon Dominated Region (PDR) and, finally, the treatment of time-dependent effects. These will will be discussed in more detail in the remainder of this section.

## 2.1 Atomic data updates

Advances in computer hardware and the construction of large scale theoretical atomic codes that can compute photoionisation cross-sections, electron-ion recombination rates, level energies, electron-collision strengths, transition probabilities and oscillator strengths for a large number of ions and transitions have allowed a comprehensive data set to be compiled. In particular, the Opacity Project and subsequently the Iron Project (Berrington et al. 1987), have produced a wealth of astrophysically useful atomic data, that is readily accessible via the TIPTOPbase website (see e.g. Mendoza et al. 2002, Nahar 2003). A review on atomic processes in PNe was given at this Symposium by M. Bautista and is included in this volume.

New calculations of total recombination coefficients of some third-row elements in ionisation stages commonly observed in PNe are currently underway (Storey, Ercolano & Badnell, in preparation). These calculations pay particular attention at the low-temperature and very-low-temperature regime and use the AUTOSTRUCTURE program of Badnell (see e.g. Badnell 1999) in the LS- or intermediate-coupling (IC) schemes, where necessary, with corrections to the theoretical level energies being applied, where experimental values are known. The coefficients are calculated for a wide temperature range and can also be applied to very cool (approx. 500 K) plasma, whose existence has been highlighted by the study of optical recombination lines (ORLs) (see e.g. the papers by Liu and Tsamis et al. published in this volume).

Observations of ORL in long-slit and integral field unit (IFU) spectra are nowadays readily obtainable. The analysis of such spectra poses some of the most intriguing questions yet to be answered in nebular astrophysics, namely the fact that ionic abundances derived from ORLs are consistently higher than those derived from the CELs in the same spectra, whilst the former diagnose temperatures that are significantly lower than those estimated using CEL ratios (He I recombination line ratios or the Balmer decrement) (see e.g. Rubin et al. 1997, Liu et al. 2000, Liu et al. 2001b, Tsamis et al. 2004). Photoionisation models of clouds containing chemical inhomogeneities and/or density discontinuities may be crucial to clarifying this long-standing problem (e.g. Péquignot et al. 2003, Tylenda 2003, Ercolano et al. 2003), however they require accurate recombination data extending to low-temperatures. Work is currently being carried out at UCL for a number of ions and some preliminary results have been presented by Bastin & Storey (2006, these proceedings).

## 2.2 Development of 3-D codes

Available computer power has increased enormously since the development of the first generation of photoionisation codes. This has allowed the construction of more complex models, including more ions, more frequency points, more lines and more atomic levels. Nevertheless, with a few exceptions, the fundamental assumption of spherical symmetry has always been retained. A glance at spatially resolved images of H II regions and PNe, that are readily obtainable with modern instrumentation, immediately demonstrates that these objects are rarely circular in projections. The effects of geometry and/or density distribution on the emerging emission line spectrum are in some cases dramatic, and an unrealistic model may result in the incorrect determination of nebular properties (e.g. elemental abundances) and/or central star parameters.

The careful combination of a number of volume-weighted spherically symmetric models (composite models) can be used to treat some aspherical geometries, such as bipolar structures (e.g. Clegg et al. 1987) or density/chemical enhancements (Péquignot et al. 2003). Morisset et al. (2005) have recently presented a pseudo-3D photoionisation code, based on this principle. The code consists of an $IDL(RSI)$ making several calls to a 1D code for a number of selected lines of sight through a 3D structure (Morisset et al., this volume). Whilst this method has the advantage of being considerably faster than running a full 3D simulation, one of its major drawbacks is that the transfer of the diffuse component of the radiation field cannot be treated self-consistently. Ercolano et al. (2003b) found appreciable discrepancies between the results obtained by a self-consistent calculation of even a simple biconical geometry and those obtained by a composite model.

The rapid development of parallel computing seen in the last few years, has made the application of stochastic techniques to the radiation transport problem an attractive option. The first 1D Monte Carlo photoionisation calculations were those of Och et al. (1998). The first fully 3D Monte Carlo photoionisation code, MOCASSIN (Ercolano et al. 2003a), uses a similar approach, but a more general version of the Monte Carlo estimators for the radiation field developed by Lucy (1999). The description of the radiation field in terms of discrete quantities allows one to simulate the individual scattering, absorption and re-emission events that characterise a photon's trajectory as it diffuses through a cloud. This public code is fully parallelised and looks to exploit the increasing accessibility of large Beowulf clusters to allow the construction of realistic nebular models.

Wood, Mathis & Ercolano (2004) also presented a 3D Monte Carlo photoionisation

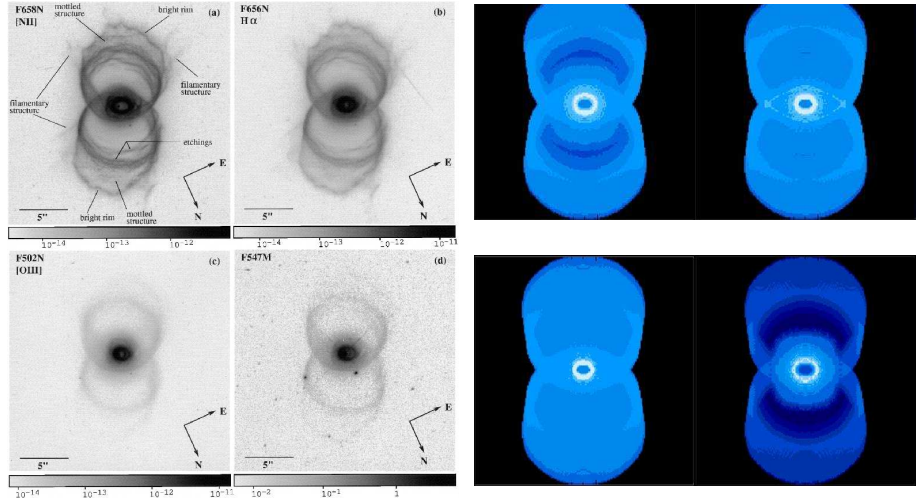

Figure 1: *Left Panel*: From Sahai et al. 1999. Narrow-band HST WFPC2 images of MyCn 18: (a) F658N ([N II]λ6586; (b) F656N (Hα); (c) F502N ([O III]λ5007); (d) a continuum filter F547N. *Right panel*: From Ercolano et al. (in prep.). 2D projections of 3D emissivity grids obtained from our best-fitting MOCASSIN model of MyCn18: (a) [N II]λ6586; (b) Hβ); (c) [O III]λ5007; (d) [O I]λ6300.

code, based on very similar techniques to those described by Och et al. (1998) and Lucy (1999), tailored for the modelling of Galactic H II regions and the percolation of ionising photons in diffuse ionised gas (Wood & Mathis 2004) .

Figure 1 (Neal, Ercolano & Sahai, in preparation) show the projected maps obtained from a model of the young PN MyCn 18, compared with the *Hubble* images presented by Sahai et al. (1999). The model images are shown to be in very good agreement with the observations, indicating that the model 3D density distribution for this object, previously presented by Dayal et al. (2000) and employed in our simulation, is a realistic description for this object. A well constrained density distribution adds confidence to the model predictions of the other nebular and stellar physical properties.

The effects of observations with a finite aperture on non-spherically symmetric objects should also be taken into account, their magnitude being larger the greater the departure from spherical symmetry of the real geometry. As an example, long-slit spectroscopic observations of extended objects, such as H II regions and PNe, are routinely performed. Whilst line ratio measurements obtained from the summed spectra of a long-slit scanned across the whole 2D projection of a nebula may be considered to be representative of the integrated nebular spectrum, they are not always available. Geometry effects on long-slit spectra at a single PA were recently studied in the case of the PN NGC 7009 (Gonçalves et al. 2006, Gonçalves et al., these proceedings) and were found to be partially responsible for the long-standing apparent overabundance of N in the FLIERs of this object (Balick et al. 1994, Hajian et al. 1997, Balick et al. 1998, and references therein).

Chemically inhomogeneous models of the hydrogen-deficient polar knots of the PN Abell 30 were presented by Ercolano et al., 2003c; these models showed that it is in fact possible to explain the extreme ORL-CEL abundance discrepancy factors (ADFs) reported for this object by a simple bi-abundance model, consisting of a dense metal-rich (cold) core of ionised gas mixed with dust grains surrounded by a less dense envelope of hydrogen deficient gas with less extreme metal abundances. In this model, heavy element ORLs, of heavy elements result are shown to be emitted virtually only by the cold core, CELs from the hotter surrounding envelope, whilst He recombination lines are emitted by both components.

In the case of some H II regions for which only mild ADFs are observed, the discrepancy may be resolved by considering temperature fluctuations in the $t^2$ prescription of Peimbert (1967). Whilst this interpretation has been successful in explaining the observations for some objects with mild ADFs, the nature and cause of the supposed temperature fluctuations remain uncertain. Even models that include density fluctuations yield only small deviations from an otherwise near homogeneous temperature structure of the emitting regions. One possibility worth investigating is whether larger fluctuations may be produced with a model including multiple ionising stars. A number of 3D MOCASSIN models are being constructed which use several geometries and ionising source distributions (Ercolano, Stasińska & Barlow, in preparation), which may help to clarify the situation from a theoretical point of view.

## 2.3 Treatment of dust grains

The importance of the effects of dust grains that are mixed with the gas in a photoionised region has long been known (e.g. Spitzer 1948), with more recent studies also aimed at investigating the role of dust grains in the thermal balance of PNe (e.g. Borkowski & Harrington 1991, Ercolano et al. 2003c, van Hoof et al. 2004). The grains are heated via the absorption of UV photons from the continuum as well as by resonance emission line photons, for example H I Ly$\alpha$, C IV $\lambda$1549, N V $\lambda$1240, C II $\lambda$1336, Si IV $\lambda$1397, Mg II $\lambda$2800. The absorbed radiation is mainly re-emitted by the grains in the IR, and this provides the main dust cooling mechanism. Additionally, the gas and dust components are coupled by a host of microphysical processes, including photoelectric emission from dust grains (for a recent discussion see Weingartner & Draine 2001), which may be an important gas-heating mechanism in H-deficient environments and in PDRs. Gas-grain collisions provide a cooling channel for the gas and a heating channel for the dust, and ionic recombination on dust grains may also have a small effect on the ionisation structure of the gas.

A new grain model was recently developed for CLOUDY and is described in detail by van Hoof et al. (2004). Furthermore, a new version (2.0) of the 3D Monte Carlo photoionisation code MOCASSIN was recently released (Ercolano et al. 2005), which includes a fully self-consistent treatment of the dust radiative transfer within the ionised region. Dust grains are allowed to compete with the ions in the gas for the absorption of UV and resonance emission line radiation, with scattering by the grains being fully implemented in the isotropic approximation or via the application of a phase function. The microphysical processes listed above, which further couple the two components, are also accounted for in the thermal and ionisation balance equations. The *average* *grain charge model* (Baldwin et al. 1991) was preferred by these authors to a fully quantised or *hybrid* (van Hoof et al., 2004) treatment in order to limit overheads, in light of the large uncertainties existing in the dust data (see Weingartner & Draine 2001, Sec. 2.3). The MOCASSIN code, which can also be run in a pure-dust mode, has the advantage of being fully three-dimensional, hence allowing the scattering problem to be treated properly. Work is currently in progress to allow polarisation maps to be produced, yielding better constraints on the geometry of the many sources for which this type of observation is available.

## 2.4 Expansion to Photon Dominated Regions

Photon Dominated Regions (PDRs – also known as *photo-dissociation regions*) are associated with all photoionised nebulae, including PNe, that are optically thick to Lyman continuum radiation. When all photons with $h\nu > 13.6\,\text{eV}$ have been absorbed H becomes predominantly neutral and this marks the beginning of the PDR (Tielens & Hollenbach 1985). Classically, the two regions have been treated as separate problems, both in terms of empirical analyses based on spectroscopic observations and in terms of the development of numerical codes that would be able to model either of the two environments. In reality, the photoionised region and PDR are closely coupled and therefore a self-consistent calculation is preferable. The radiation field impinging on the edge of a PDR is, in fact, a result of photon diffusion through the ionised region, and is naturally calculated by a common photoionisation code, whilst classical PDR codes generally treat $G_0$ as a free parameter to be determined from the model. Moreover, some of the most important IR diagnostics used to characterise the gas in a PDR are also partially emitted in the ionised region. This is the case, for example, for all transitions of ions such as $O^0$, $C^+$, $Si^+$, having ionisation potentials (IPs) lower than 13.6 eV.

Many PNe are known to be surrounded by extended PDRs, as is the case, for example, of NGC 7027 (e.g. Graham et al. 1992, Latter et al. 2000) and NGC 6302 (e.g. Liu et al. 2001a). More recently, PDRs have also been detected around cometary knots, such as those of the Helix nebula (O'Dell, Henney & Ferland 2005; Tarter et al. and Manchado et al., this volume ) and the Ring nebula (Speck et al. 2003).

A new version of CLOUDY was recently released, which allows self-consistent calculations of the spectrum, chemistry, and structure of photoionised clouds and their associated PDRs to be performed (Abel et al. 2005). Besides the improvements to the grain model mentioned in the previous section, a molecular network was also included (Abel et al. 2005) with approximately 1000 reactions involving 68 molecules. Recent advances in the treatment of $H_2$ are described by Shaw et al. (2005). The chemical equilibrium was benchmarked at the Leiden 2004 PDR workshop (http://hera.ph1.uni-koeln.de/r̃oellig/, Roellig et al. in prep.) and found to be in agreement with the results of other PDR codes.

Having completed the inclusion of dust grains and subsequent benchmarking phase, the MOCASSIN code is also currently undergoing further development, with the incorporation of the molecular reaction network in use by the UCL_PDR code (Bell et al. 2005, and references therein), which is also one of the codes recently benchmarked at the Leiden 2004 PDR workshop. The new version of the MOCASSIN code, which will be described in a forthcoming paper by Ercolano et al. (in preparation), will allow to make self-consistent models of ionised regions and PDRs for arbitrary geometry and density distributions, including, for example, cometary knots and their shadow regions.

## 2.5 Time-dependent effects

The ionisation structure and electron temperature of a photoionised cloud is obtained by simultaneously solving the thermal balance and ionisation equilibrium equations, classically under the steady-state assumption. This approximation is valid when the atomic physics timescales are short compared to those of gas-motions or the rate of change of the ionising field. If this is true the photoionisation problem can be considered to be fully decoupled from the dynamics; in this case a given gas density distribution represents a *snapshot* of the cloud evolution and a time-independent photoionisation simulation can be performed. PNe and H II region can generally be expected to be mostly in equilibrium, with dynamical effects only becoming significant in a very small region near the ionisation front, hence having near-negligible effects on the integrated emission line spectrum (e.g. Harrington 1977).

The cases studied by Harrington (1977) were limited to weak D-fronts with subsonic gas motions with respect to the front. There are, however, a number of instances when dynamics may have substantial effects on the predicted spectrum. Transonic photo-evaporation flows, such as those believed to be occurring in the cometary knots of PNe (López-Martín et al. 2001), cannot be accounted for under the assumption of thermal and ionisation equilibrium.

In cases where shocks are present the gas cannot be considered to be in equilibrium, and classical time-steady codes are not appropriate. The *mappings* III code (Sutherland & Dopita 1993) was designed to deal with shock ionisation and has been applied to date to a variety of astrophysical environment.

Time-steady dynamics has recently been added to CLOUDY; the numerical algorithms employed and a set of self-consistent dynamic models of steady ionisation fronts are presented in recent papers by Henney et al., (2005a,b).

## 3 Conclusion: The Future

It has been the aim of this review to identify and discuss in broad terms the most recent advances in photoionisation codes. Several large scale codes exist and are being maintained by individual working groups, coming together regularly to ensure the reliability of the results via the running of a set of benchmark problems. Modern codes aim at the construction of realistic simulations in order to efficiently extract information from a wealth of new spectroscopic observations available over large wavelength ranges and for increasingly more distant and faint objects.

The coupling of photoionisation/PDR and hydrodynamic RT codes is only a step away. This is perhaps one of the most exciting prospects, as the new simulations will undoubtedly yield new insights into the physics of many astronomical phenomena. Already some photoionisation codes allow for time-dependent effects to be taken into account. In the near future it will be possible to perform a full photoionisation/PDR calculation at each time-step in a large hydrodynamic simulation, to follow, for example, the evolution of the time-dependent non-equilibrium spectrum of a dynamically evolving gas cloud, from the time it starts being illuminated. Both the physics and numerical techniques are mature, it is only a matter of waiting for the required computer power to become available.